# Light Element Abundance Inhomogeneities in Globular Clusters: Probing Star Formation and Evolution in the Early Milky Way

Michael M. Briley
McDonald Observatory, University of Texas
Austin, TX 78712-1104 USA

Roger A. Bell
Department of Astronomy, University of Maryland
College Park, MD 20742-2421 USA

James E. Hesser
Dominion Astrophysical Observatory
Herzberg Institute of Astrophysics
National Research Council
5071 W. Saanich Rd., RR 5
Victoria, B.C. V8X 4M6

and

Graeme H. Smith
Lick Observatory, University of California
Santa Cruz, CA 95064 USA

Direct correspondence to Dr. Michael Briley, (512) 471-0500, (512) 471-6016(fax)
Internet: mike@maxwell.as.utexas.edu

## ABSTRACT

Abundance patterns of the elements C, N, and O are sensitive probes of stellar nucleosynthesis processes and, in addition, O abundances are an important input for stellar age determinations. Understanding the nature of the observed distribution of these elements is key to constraining protogalactic star formation history. Patterns deduced from low-resolution spectroscopy of the CN, CH, NH, and CO molecules for low-mass stars in their core-hydrogen or first shell-hydrogen burning phases in the oldest ensembles known, the Galactic globular star clusters, are reviewed. New results for faint stars in NGC 104 (47 Tuc, C0021-723) reveal that the bimodal, anticorrelated pattern of CN and CH strengths found among luminous evolved stars is also present in stars nearing the end of their main-sequence lifetimes. In the absence of known mechanisms to mix newly synthesized elements from the interior to the observable surface layers of such unevolved stars, those particular inhomogeneities imply that the original material from which the stars formed some 15 billion years ago was chemically inhomogeneous in the C and N elements. However, in other clusters, observations of abundance ratios and C isotope ratios suggest that alterations to surface chemical compositions are produced as stars evolve from the main sequence through the red giant branch. Thus, the current observed distributions of C, N, and O among the brightest stars (those also observed most often) may not reflect the true distribution from which the protocluster cloud formed. The picture which is emerging of the C, N and O abundance patterns within globular clusters may be one which requires a complicated combination of stellar evolutionary and primordial effects for its explanation.

astro-ph/9408007   2 Aug 94

# 1. Gerhard Herzberg, Diatomic Molecules and The Birth of the Galaxy

Throughout his distinguished career, Dr. Herzberg's focus on the power of molecular spectroscopy to resolve fundamental astrophysical problems has been a recurrent theme. Indeed, the final 15 pages of his renowned monograph, Molecular Spectra and Molecular Structure I. Spectra of Diatomic Molecules [1], are devoted to astrophysical applications[1]. In the near-ultraviolet to blue region of the spectra of cooler stars, absorption band features arising from the CH and CN molecules are often prominent and have played critical roles in the empirical classification and physical interpretation of stellar spectra. In the past two decades, such features have been found to offer intriguing clues to the evolution of low-mass stars and star formation in the halo of the Milky Way galaxy some 15 billion years ago. In this brief review, we present background on the astrophysical problem and discuss recent observations of CN, CH, NH, and CO bands that reveal important diagnostic patterns in the oldest known stars of the Galaxy.

Superficially, astronomers' most basic picture of the structure of the Milky Way galaxy has changed little since the seminal analysis by Plaskett and Pearce in the 1930s [2, 3] confirmed the hypothesis proposed by Lindblad and Oort [4, 5] that the Milky Way is a flattened, differentially rotating galaxy. Our detailed understanding, however, has been vastly enriched during the following decades (for non-technical reviews see, e.g., [6, 7]). As viewed today, the Galaxy consists of a huge rotating disk, containing ~98% of the visible mass, within a larger, more slowly rotating halo containing the remaining luminous mass. Of the latter component, we focus here on detailed studies of a few of the 100 or so globular star clusters that are members of the Galactic halo, although these clusters contain in total only about 2% of all halo stars. Stars appear to constitute but a fraction of the total mass of the halo, much of which is comprised of the mysterious dark matter.

Through comparisons between the observed properties of the cluster stars and theoretical stellar models, the globular star clusters offer unique opportunities to measure ages, and hence establish relative and absolute chronologies, for the oldest constituents of the Galaxy. Moreover, they also provide the most direct estimate of the age of the Universe.

---

[1] Indeed, claiming the Universe as our laboratory, astronomers and astrophysicists would maintain that most, if not all, of his Chapter VIII is devoted, in one way or another, to astrophysical applications.

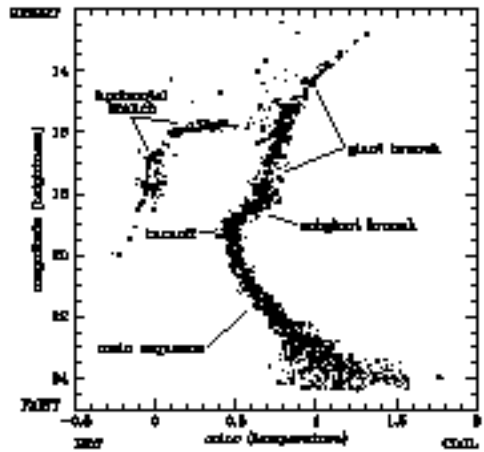

FIG. 1: Observed luminosities in the Johnson Visual photometric system (ordinate) and Johnson Blue minus Visual colors (abscissa) for stars in the globular cluster M15 (courtesy P. B. Stetson; the original data are from [78]). Phases of stellar evolution referred to in the text have been identified. The location of the turnoff (the end of the main-sequence phase) is dependent on the age and chemical makeup of the cluster stars (the more massive stars having exhausted the H in their cores and evolved into giants). By comparing the morphology of this diagram - and, in particular, the luminosity of the turnoff - with theoretical models, the age of the cluster can be determined.

For the past fifteen years, astronomers have been evaluating two extreme views of how the Milky Way galaxy might have formed. Eggen, Lynden-Bell and Sandage [8]; Sandage [9], postulated that perhaps a single massive protogalactic cloud underwent a rapid collapse (in a few $\times 10^8$ years), during which time the chemical composition rapidly built up from nearly primordial element abundances to those typical in the disk of the Galaxy today. In 1978 Searle and Zinn [10] proposed an alternative scenario in which the Galaxy formed by chaotic mergers of fragments, within which independent chemical evolution occurred. The real story is doubtless more complex than either of these scenarios; we note that, at present, arguments for merger processes having played a significant role in the formation of the Galaxy and its halo are in ascendancy. To map the formation chronology of the Galaxy is a daunting task; a key element of that task is determining accurate ages for the various major constituents, especially the oldest ones capable of yielding insight into processes in the protogalaxy.

Determination of the ages of individual stars in the distant halo or bulge of the Galaxy is extremely difficult, in part because of uncertainties in their distances, and hence in their derived total energy outputs (which is a key property for comparison with stellar models). Distances to star clusters, on the other hand, can be much more reliably estimated, and relative luminosities and temperatures can be determined to high precision from observational data. Moreover, in such clusters, copious numbers of coeval, equidistant stars are found in stages of evolution which depend, to first order, only on the mass of each star (see Figure 1). By eliminating evolutionary and other ambiguities common in the study of field stars, the comparison of cluster stars with theoretical stellar models is facilitated (e.g. [11]).

Such studies show that, in general, the system of open clusters located in the disk of the Galaxy exhibits a wide range of ages (from $\approx 10^5$ to $\approx 10^{10}$ yrs) and a relatively small range of abundances for elements heavier than He (less than about a factor of four in the ratio of heavy elements to H)[2]. On the other hand, the globular cluster system exhibits a range in heavy element content of a factor of 100 (with some clusters in the Galactic bulge having compositions almost as enriched, relative to primordial values, as those found in solar neighborhood objects), and a range of ages extending over $\approx (12\text{-}16) \times 10^9$ years, with an uncertain distribution that may peak around the older ages and include but a small number of younger clusters (for a recent review, see [12]).

The globular cluster system of the Galaxy has at least two kinematically separate subsystems [13-16]. One, generally referred to as the halo subsystem, rotates at less than or about 25% of the speed of the disk and has an average chemical composition of heavier elements which is smaller by a factor of 40 than that of the Sun. A second, often referred to as the thick-disk subsystem, rotates at a similar speed to the Galactic disk and is deficient in heavy elements by an average factor of 3 with respect to the Sun. The relation between these and other possible subsystems is a matter of active research today.

In part, that intense interest stems from the fact that our knowledge of the age of the Galaxy is set by the ages of the oldest Galactic globular clusters. One of the most important factors which enters into the reliability of their age determinations is the chemical composition of the stars therein, as composition governs the nuclear reaction rates and atmospheric opacities, which in turn govern the stellar energy outputs [17]. The chemical composition of stars depends upon when and where they formed. The earliest nucleosynthesis of elements with atomic weights greater than that of boron is expected to have occurred in stars some 10 - 100 times more massive than the Sun [18], all of which are relatively short lived. As the first Galactic generation of such stars exploded in violent supernova events, it enriched the interstellar gas (from which subsequent generations of stars formed) with elements heavier than He that had been synthesized both within the stellar cores and during the supernova explosions. Lower-mass stars may also have contributed heavy elements to the halo at a somewhat later epoch, e.g., supernovae of Type I, which are thought to result from accreting white dwarf stars in binary systems, can produce substantial quantities of iron-peak elements [18], while significant quantities of heavy s-process elements such as barium are thought to be nucleosynthesized within stars of mass less than 10 $M_{sun}$ [18]. Different types of stars and supernovae produce different relative abundance ratios. Thus, studies of stellar chemical compositions probe the integrated history of the formation of chemical elements in the Galaxy by differing generations of stars.

With this abbreviated introduction, we turn now to the specific problem with which we and many others have been wrestling for two decades: a problem in which Dr. Herzberg's small molecules have played a seminal diagnostic role.

## 2. Globular Cluster Abundances: The Overall Nature and a Brief History of the Problem

Within a given globular cluster (with only two known exceptions), the chemical composition of the stars is extremely uniform in elements heavier than silicon (as reviewed in [12]). Globular clusters contain many cool stars in whose atmospheres simple diatomic molecules and radicals such as CN, CH, CO, and NH are able to form. The spectra of such stars exhibit absorption bands due to these species.

Some twenty years ago, analyses of intermediate-band photometry [19] and low-resolution spectra [20-22] detected differences in the strengths of several absorption bands arising from CN and CH among a small number of the higher luminosity red giant stars in a few clusters. This was the first compelling indication that individual globular clusters were not completely chemically homogeneous. A subsequent and more extensive photometric survey [23] showed that variations of CN band strengths (both the $\Delta v = 0$ and $\Delta v = -1$ sequences of the B $^2\Sigma^{+-}$ X $^2\Sigma^+$ transition, with bandheads at 388.3nm and 421.6nm, respectively) were ubiquitous within the eight clusters of those surveyed whose stars had a sufficiently high overall chemical composition[3] (i.e., [M/H] > –1.6) for the band strength variations to be measurable by the photometric technique employed. Stars of four other clusters with [M/H] < –1.6 showed no convincing differences, as absorption from these bands was too weak to be detectable by this technique. Since the abundance of the CN molecule depends upon the product of the abundances of two "heavy elements," the CN abundance is very sensitive to the overall heavy element abundance within a stellar atmosphere, and consequently the CN bands weaken considerably with decreasing metal abundance. Hesser et al. [23] suggested that the behavior of the CN bands among stars of the halo globular cluster population was indicative of a fundamental difference in the properties of the CNO elements between these cluster stars and the Galactic disk. In fact, it was subsequently found by Kraft et al. [24] that the pattern of C and N abundances among the globular cluster stars was even dissimilar to that evinced by the non-cluster stars in the Galactic halo.

---

[2] For historical reasons, such elements are refered to as "metals" by astronomers: a terminology used herein.

Evidence soon emerged that stars within some clusters of overall composition [M/H] > –1.6 exhibited variations in CN strengths that were distributed in a bimodal pattern [25]; reviews can be found in [26-28]. In addition, a number of studies of evolved red giants in many of the "bimodal-CN" clusters (e.g., Da Costa and Cottrell [29] and Norris et al. [30] for the cluster NGC 6752) found an anticorrelation between the strengths of the 430.0nm "G band" of the CH molecule (A $^2\Delta$ - X $^2\Pi$) and the CN bands. A certain predilection has also been found for CN-strong stars to be more numerous than CN-weak stars in some clusters, while CN-weak stars are more numerous in others. The relative numbers of CN-strong to CN-weak stars does not appear to be correlated with other cluster characteristics, such as overall chemical composition [M/H] and horizontal-branch morphology[4], except for the puzzling observation of Norris [31] that the percentage of CN-strong giants appears to be roughly correlated with the degree of ellipticity in the shape of a cluster.

---
[3] [M/H] = $\log_{10}$ (M/H)$_{star}$ - $\log_{10}$(M/H)$_{solar}$. The letter "M" is used here to refer to abundances of elements heavier than He. In the literature the iron abundance [Fe/H] has often been used as a proxy for most heavy elements, although it is now evident that the abundances of a number of metals do not exhibit a consistent ratio to that of Fe. Generally, then, globular cluster stars are much more deficient in heavier elements than the Sun. Since, according to current estimates, the Sun formed some 7 - 11 × $10^9$ years after the halo globular clusters, it follows that the composition of the Sun has been determined by many more generations of stellar nucleosynthesis, i.e., the Sun formed at a time when the build-up of heavy elements within the Galaxy had progressed far beyond that at the epoch of Galactic globular cluster formation.
[4] The incidence of CN-strong stars is therefore not strongly coupled to the long-standing "Second Parameter" problem, whereby clusters of similar overall metallicities are observed

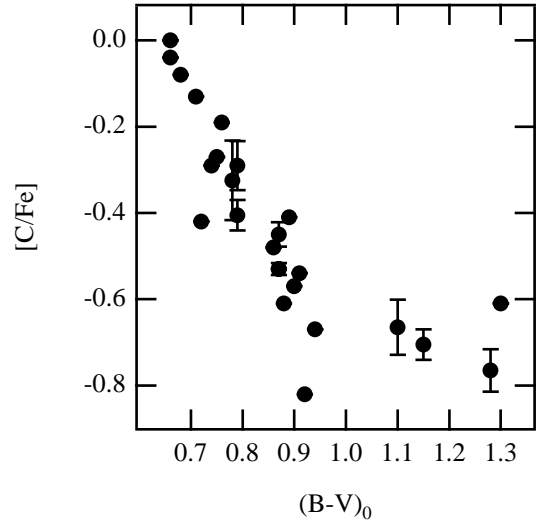

FIG. 2: C abundances measured as a function of evolutionary state for RGB stars in the metal-poor cluster NGC 6397 (from [39]). The stars evolve from left to right (becoming redder) while also becoming increasingly more C deficient.

The more metal-poor clusters (with [M/H] < –1.6) yielded a somewhat different pattern of inhomogeneities. A careful examination of CH band strengths in three such clusters (M92, NGC 6397, and M15) by Bell et al. [32], Carbon et al. [33], and Trefzger et al. [34] revealed a trend of decreasing carbon abundance with increasing luminosity among stars in the red giant branch (RGB) phase of evolution (Figure 2). Such observations imply that these metal-poor cluster stars, which are convective throughout much of their extended envelopes, are somehow able to mix to their surface material that is relatively deficient in carbon. In addition, this mixing appears to be continuously active throughout much of the RGB phase of evolution.

When considered in concert with the anticorrelated CN and CH abundances found in the more metal-rich clusters, these observations seem consistent with the following explanation. After a star consumes the hydrogen in its core, it evolves into a red giant, with energy being supplied by a hydrogen burning shell. The p-p chain of hydrogen fusion in this shell is accompanied by CN-cycle (Bethe) processing at the very outside of the shell, in which $^{12}C$ is used as a catalyst to convert H into He. The rates of the CN-

---
to have generally different color distributions among their horizontal branch (core He-burning) stars.

cycle chain are such that the $^{14}$N$(p, \gamma)^{16}$O reaction runs the slowest and effectively converts all of the $^{12}$C into $^{14}$N (98% $^{14}$N at equilibrium). Material processed in this way will become C-poor and N-rich. The solar ratio of C to N is 4.2:1. Therefore, in a star with a solar C to N ratio, the abundance of free N regulates the formation of CN (N is the minority species). Suppose, however, that material within the envelope and atmosphere of a globular cluster giant is circulated through an interior region of the star within which the CN-cycle processing is taking place on a sufficiently short timescale to allow the transformation of $^{12}$C into $^{14}$N. This material is eventually returned to the stellar surface. As a consequence, the resulting C:N abundance ratio in the atmosphere will be decreased, and a large CN abundance will be produced as a result of the large percentage increase in the $^{14}$N abundance; the CN bands will become stronger [27, 35]. A maximum CN band strength will then be achieved when $^{12}$C to $^{14}$N processing has decreased the C:N number abundance ratio to approximately unity (only approximately unity because of the effects of CO formation [35]). Once the ratio falls below unity, carbon becomes the minority species controlling CN formation, so that any further decrease in the surface carbon abundance as a result of continued internal CN-cycle processing will bring about a decrease in the CN abundance. As such processing proceeds, the abundance of free carbon will continuously decrease, and the CH bands will also become weaker.

The observed decline of carbon abundance with luminosity on the giant branches of some clusters [33, 34, 36-39] suggests that somehow the C and N abundances in the atmospheres of cluster giants are being continuously modified via CN-cycle processing throughout much of their red giant branch phase of evolution. That the degree of such processing can become very extreme by the time that globular cluster giants have evolved to relatively high luminosities is suggested by measurements of the nitrogen abundances of such stars in the clusters M92, M3, M13, and M15 by Carbon et al. [33], Suntzeff [36, 37], and Trefzger et al. [34] from analyses of the 336.0nm NH absorption bands in their spectra. These studies find that although the carbon abundance decreases with increasing luminosity amongst the red giants, the nitrogen abundance remains essentially constant on average. Such a situation can occur if a large percentage of carbon has already been converted to nitrogen, such that any further carbon conversion will bring about only a small fractional increase in the nitrogen abundance [33].

A necessary consequence of the presence of extensively CN-cyled material in the envelope of a red giant (or any star) is a very low $^{12}$C/$^{13}$C ratio [40]. For material that has been processed to an equilibrium chemical composition through the CN-cycle the value of the $^{12}$C/$^{13}$C number-abundance ratio is about 3, while the solar ratio is approximately 90. Therefore, as the envelope of an evolving star is progressively cycled through a region of CN-processing, the $^{12}$C/$^{13}$C ratio is expected to drop precipitously [41].

One of the first applications of high-resolution astronomical infrared spectrographs to globular cluster studies was to examine the 2.35 μm vibration-rotation bands of the CO molecule. The advantage of observing in the infrared in this case is the large isotopic shift between the $^{12}$CO and $^{13}$CO bandheads at 2.295 μm and 2.345 μm, respectively; the relative strengths of the $^{12}$CO and $^{13}$CO absorption features can be used to yield $^{12}$C/$^{13}$C ratios. Observations of the brightest red giants in M4 and M22 by Smith and Suntzeff [42], 47 Tucanae by Bell, Briley, and Smith [43], and in NGC 6752 and 47 Tucanae by Suntzeff and Smith [44] revealed that, indeed, the $^{12}$C/$^{13}$C ratios were quite small ($3 < {}^{12}$C/$^{13}$C $< 10$), regardless of whether a star was CN-strong or CN-weak. Similar studies which use the 800.2nm $^{12}$CN and 800.5nm $^{13}$CN features in the spectra of stars in M4, M22, Omega Cen, 47 Tucanae, and M13 have been carried out by Brown and Wallerstein [45], and Brown, Wallerstein, and Oke [46], and most recently in M71 by Briley, Smith, and Lambert [47], with the same result: very low $^{12}$C/$^{13}$C ratios for every cluster red giant that was observed.

That the carbon isotope ratios are so low for all stars, regardless of CN band strength, implies that the CN-weak stars, which might have been expected to reflect the initial distribution of C and N abundances, must have also had their envelopes at least partially cycled through a region of CN-processing. Perhaps even more curious, among the more metal-poor clusters ([M/H] < –1) for which $^{12}$C/$^{13}$C ratios have been measured, there is no difference in this ratio between CN-strong and CN-weak stars. Yet, as discussed by Briley, Smith, and Lambert [47], this appears not to be the case among the more metal-rich stars, where the CN-weak stars have been shown to have slightly higher $^{12}$C/$^{13}$C ratios than their CN-strong counterparts.

Within the framework of a stellar-interior-mixing hypothesis, is it possible to explain, at least qualitatively, the CN variations, the CN versus CH anticorrelations, the low $^{12}$C/$^{13}$C ratios, and the decline of C abundances as stars of metal-poor clusters progress through the red giant phase of their evolution? Canonical models of globular cluster stars indicate that throughout the red giant phase of evolution, the base of the convective envelope is not in contact with regions of active CN-cycle processing. If such models are correct, then the observations imply that some other form of mixing must connect the base of the envelope with the actively processing regions. A candidate mechanism for mixing within the deep interior, namely mixing via meridional circulation currents, has been put forth by Sweigart and Mengel [48]. As outlined by them, the principal barrier to such mixing is a molecular weight gradient which exists within lower luminosity giants between the base of the convective envelope and the region of CN-cycling. However, the H-burning shell in an evolving red giant moves outwards as the star evolves to higher luminosities, and this gradient is calculated to be eventually destroyed, after which meridional circulation currents can then cycle envelope material through the CN-processing region at the outer edge of the energy generating shell. It is of interest to note that the destruction of the molecular weight gradient within very metal-poor stars is calculated to occur at a luminosity similar to that of the faintest stars for which carbon depletions were reported by Bell et al. [32], although later spectroscopic observations by Langer et al. [38] suggest that the C depletions begin among stars of even lower luminosity.

During the RGB phase of evolution, the initial molecular weight gradient (which prohibits mixing) and the extent of the zone of CN-processing beyond the main hydrogen-burning shell both depend on the metallicity of the star. This gradient is smaller and the zone of CN-processing is thicker in a metal-poor giant than in a metal-rich giant of the same luminosity. Thus, circulation of the envelope through the zone of CN-cycle processing should be easier in a metal-poor star. Indeed, observations of smaller carbon depletions among cluster red giants with larger overall metal abundances have been reported by a number of investigators [42, 49, 50].

Consistent with the expectations of more effective deep mixing within the lower-metallicity cluster giants, the value of the $^{12}C/^{13}C$ ratio in all stars measured to date in the metal-poorest clusters is found to be near the equilibrium value; under such extreme circumstances any additional mixing of the CN-strongest stars would not result in an appreciable decrease in their $^{12}C/^{13}C$ ratios. However, among the red giants in the more metal-rich clusters 47 Tuc and M71, significant differences in $^{12}C/^{13}C$ between the CN-weak and CN-strong stars have been observed (see Figure 2 of [47]). This result is also consistent with there being a metallicity dependence in the effectiveness of the deep-mixing mechanism hypothesized to be responsible for these extreme isotope ratios.

That other mixing processes in addition to convection may be occurring within the deep interiors of globular cluster red giants has significant implications for theories of the evolution and structure of low-mass stars (which traditionally do not include such processes) and for the use of globular clusters for Galactic age determinations. Langer and Kraft [51] rigorously explored the issue of whether the cluster CN, CH, and NH inhomogeneities known to them at the time could be explained entirely by the transformation of material in the atmospheres of cluster stars by the CN-process only. Such processing would not necessarily require mixing to penetrate deeply into the CN-cycling zone above the main hydrogen burning shell. However, recent studies of the 630.0nm OI atomic line and CO molecular band strengths among red giants in a number of clusters (NGC 6397, M92, M15, M3, M13, M5, M71) by Pilachowski, Sneden, Kraft, Langer, Bell, Brown, Wallerstein, and others [46, 52-59], have changed this picture considerably. These researchers find that CN enhancements are often accompanied by surface oxygen abundance depletions, a result which suggests that RGB mixing may extend so deep as to also bring O-poor/N-rich ON-cycled material up into the stellar envelope (from an area just below the region of CN-cycling). If this is the case, then the CNO abundances which have been determined from bright cluster red giants may not reflect the original abundance distribution of the gas from which a cluster formed. Yet oxygen is the third most abundant element in normal stars and a principal source of opacity in stellar interiors. An error in [O/Fe] of 0.35 dex can affect ages determined from globular cluster color-magnitude diagrams by ~ 1.7 Gyr [17].

Out of necessity, most of the earliest studies of light element abundance inhomogeneities among globular cluster stars focused on the brightest, and therefore among the most evolved, of the cluster giants. The interpretation of these inhomogeneities as evidence that processes at work within the observed cluster giants themselves had transformed the original element abundance ratios (albeit by variable amounts) seemed quite natural, especially when such an interpretation seemed to account for so many of the observations. But even before the introduction of highly sensitive CCD detectors for use with astronomical spectrographs, there were a number of other observations which could not be readily explained through processes internal to the cluster stars themselves.

As early as 1978, it was noted by Hesser [60], Hesser and Bell [61], and Bell, Hesser, and Cannon [62], that significant star-to-star differences in 388.3nm CN band absorption existed among less evolved stars in the nearby cluster 47 Tucanae. Some of the stars observed by them are sufficiently faint as to be in the main-sequence stage of evolution, within which CN-processing is less vigorous and is confined to a hydrogen-burning core. In addition, such stars have convection zones which extend to much shallower depths below the atmosphere than do red giants.

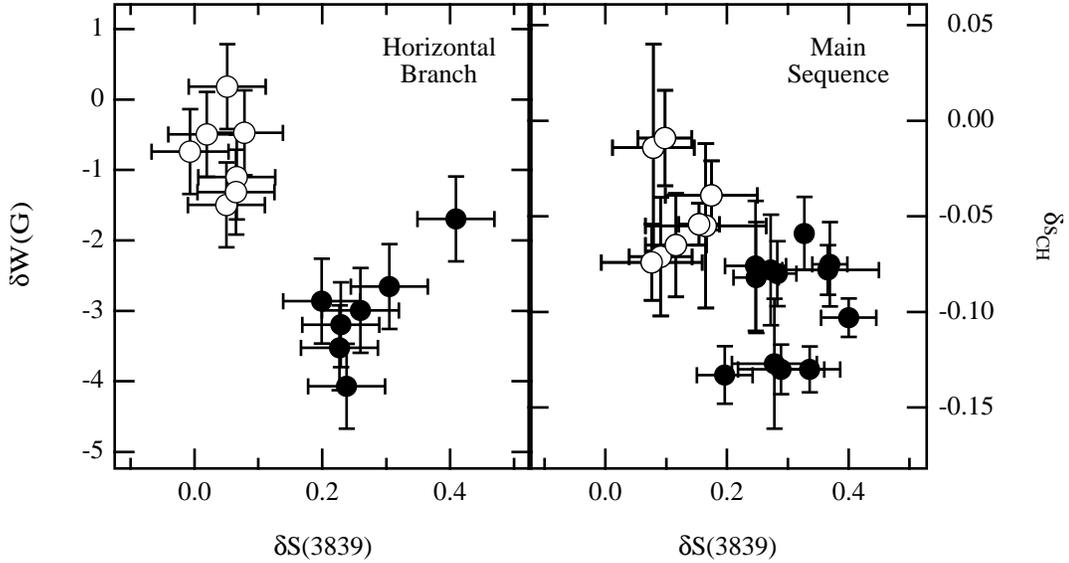

FIG. 3: CN ($\delta S(3839)$) and CH ($\delta W(G)$ and $\delta s_{CH}$) band strength excesses measured from evolved horizontal branch and unevolved main-sequence stars in 47 Tucanae (from [64, 66], large values indicate stronger absorption). The similarity in the pattern of abundance inhomogeneities between the two groups is striking, despite the considerable difference in their evolutionary states (note that the two studies used different CH sensitive indices).

Following the initial studies of the main-sequence (unevolved) stars of 47 Tucanae, little additional work was done in this area until the end of the 1980's because of the prohibitive integration times required to observe these faint stars, even on the largest telescopes. However, with the introduction of sensitive CCD detectors and multiobject spectrographs, investigators have recently returned to these stars.

In 1989, Suntzeff [37] reported the existence of 388.3nm CN variations among several main-sequence stars in NGC 6752. Also present was a general anticorrelation between CN and CH band strengths. A similar result was observed among a sample of ten main-sequence stars in 47 Tucanae by Briley, Hesser, and Bell [63]. The program stars in both these samples are undergoing core hydrogen burning and have yet to begin their red giant evolution. That CN band strength differences are observed among these stars suggested that the original material from which the clusters formed some 15 billion years ago was chemically inhomogeneous in the elements carbon and/or nitrogen. Moreover, the patterns of abundances were notably similar to those observed among the more evolved red giant stars. Indeed, the abundance analysis by Briley, Hesser, and Bell [63] implied little change in the range of carbon and nitrogen abundances with evolution. However, both of these studies [37] and [63] suffered from either small sample sizes or samples biased towards the extrema in molecular band strengths, so that only limited conclusions could be drawn regarding the overall distribution of C and N abundances on the main-sequence.

Recently, Briley et al. [64], studied twenty randomly selected main-sequence stars in 47 Tucanae in order to better determine the distribution of CN and CH band strengths. They report a significant star-to-star spread in their measured CN band index. The ratio of CN- strong to CN-weak stars is 12:8, consistent with that found among the more luminous red giants [65] and among the even more highly evolved horizontal-branch stars [66] (the ratio on both branches is slightly greater than 1:1). Moreover, a CN versus CH anticorrelation was also found among this sample of main-sequence stars. Similar anticorrelations have been observed among the brighter cluster members in 47 Tucanae [67], and many other clusters, as well (and have long been thought to be a signature of mixing processes). Yet the 47 Tucanae main-sequence stars exhibit C and N abundance characteristics similar to those of their more evolved counterparts (see Figure 3). An analysis with synthetic spectra suggests that the range of C and N abundances is consistent with that found among the brighter stars. These results indicate that at least some component of the CN and CH variations among the bright stars is likely the result of composition differences present when the stars formed [64].

Another surprising result which initially seemed to support the concept of primordial abundance variations was the documentation by Peterson [68], Cottrell and Da Costa [69], Norris et al. [30], Norris and Pilachowski [70], Drake, Smith, and Suntzeff [71], and others (*e.g.*, [56, 58]), of correlations between CN band strengths and the abundances of Al and Na among stars in a number of clusters. Nucleosynthesis calculations indicated that Na and Al can be produced by nuclear reactions associated with hydrostatic carbon-burning and neon-burning within massive stars whose central temperatures reach $0.8 - 1.2 \times 10^9$ and $1.3 - 1.7 \times 10^9$ K, respectively [18]. However, the present-day cluster giants are of such low mass that their interior temperatures are not expected to get high enough to support carbon and neon fusion. Consequently, the patterns of Al and Na enhancements were thought to have been created by a process exterior to the present cluster stars [69]. Moreover, the anticorrelation of C to N abundances observed in the metal-poor clusters by Carbon et al. [33] and Trefzger et al. [34] is not strictly one-to-one, as would be expected if the N excesses came entirely from the CN-cycle. They reasoned that there is probably an underlying initial C and/or N inhomogeneity which could not be attributed to mixing. All of these observations are significant barriers to the idea of interior or mixing processes being responsible for all of the intracluster inhomogeneities in the abundances of elements lighter than silicon.

Taking all of the observations into account appears to result in a dilemma: different results support one scenario or another, and it has not yet been possible to find a single unifying explanation for all of the data. The declining C abundances with RGB luminosity are clearly the result of an evolutionary (or mixing) process. This is also consistent with the observation that, on average, the CN band strengths anticorrelate with CH. Yet the presence of CN inhomogeneities among unevolved stars, the lack of strict C/N anticorrelations, and the Al and Na anticorrelations with CN, have all been used to argue that the origin of the variations cannot be internal to the present day stars, and that the C and N abundance patterns must therefore have been fixed very early in the history of the stars. Problems with deep mixing have also been noted by some authors with regard to the effects that it would have on other aspects of stellar evolution [40, 41, 72].

## 3. Discussion

Recent work, particularly with regard to O, Na, and Al, that has built on the initial discoveries of Peterson [68], Cottrell and Da Costa [69], Norris et al. [30], and Pilachowski [52], has produced new insights into the origin of chemical element inhomogeneities within globular clusters. The observations of large (> 0.6 dex) O abundance variations among bright giant stars [46, 53, 55, 56, 58, 59] in clusters with narrow main-sequence turn-offs implies that these variations must be the result of some process taking place during the red giant phase of evolution. If main-sequence stars also possessed such large differences in oxygen abundance then they would evolve at significantly different rates (owing to differences in interior opacities), such as to cause a larger temperature spread in the turnoff region of the color-magnitude diagram than is observed [17]. This constraint, when coupled with the progressive C depletions observed along the red giant branches of metal-poor clusters, argues for the occurrence of a mechanism which can cycle the envelopes of cluster giants not only through a region of CN-processing, but also of ON-processing, during RGB ascent.

This picture has been given added feasibility by the recent work of Langer, Hoffman, and Sneden [73], who have demonstrated that it may also be possible to produce Na and Al through proton captures in the ON-processing shell of a red giant. This then could explain the Al and Na versus CN correlations, as well as the Na versus O anticorrelations, observed among cluster RGB stars (*e.g.*, [55, 56, 58, 59]) without the need for a primordial source. The tendency for the metal-richer clusters such as M4 and M71 to evince smaller star-to-star Na variations than the more metal-poor ones (such as M13) might then be accounted for by the proposal (noted above) that there is a metallicity dependence to the mixing efficiency.

However, serious questions still remain about the viability of such a mechanism for producing the observed CN variations. One of the most difficult to understand is why pronounced CN variations have not been observed among Population II subgiant stars in the field. If the CN inhomogeneities are produced through some process which is internal to metal-poor stars, what is it about the cluster environment that enables it to operate? Since the outer convective region of main-sequence turn-off stars is very thin and contains little mass (compared to the deep convective envelopes of RGB stars), small amounts of material accreted onto the atmospheres of such stars during an earlier epoch of their main-sequence evolution could affect their spectra in detectable ways. (Such accretion might occur as the star orbits within the cluster potential well, where some stellar ejecta might accumulate, or as the cluster plunges through the Galactic plane). However, the data for 47 Tucanae stars imply that their CN, CH behavior did not arise from an accretion process during their core-hydrogen burning main-sequence phase. If this were the case, the enriched outer layers would have become diluted with unmodified material from the interior when the depth of the convective layer increased as the stars evolved into red giants, contrary to the observed similarity of the C and N inhomogeneities between the main sequence turn-off and red giant branch stars in 47 Tuc.

That CN inhomogeneities have been observed among main-sequence stars of at least two globular clusters argues strongly that at least some component arose from a primordial source; *i.e.*, the cluster stars formed some 15 Gyr ago from material that had already been differentially enriched by more massive stars. A scenario which might accord with such considerations is one similar to that of Cottrell and Da Costa [69] in which many of the low-mass stars in a globular cluster were still in a pre-main-sequence phase of evolution at a time when a number of the more massive (~ 5 $M_{sun}$) stars ejected their N-rich, C-poor outer envelopes back into the interstellar medium via stellar winds (relatively quiescent mass loss via stellar winds would not disrupt a globular cluster in the way that a large number of supernova explosions might [74, 75]). The ejected stellar envelopes might have had all the signatures of the CN, and possibly the ON-cycle, including low $^{12}C/^{13}C$ ratios. Consequently, if this material was accreted onto, or otherwise incorporated into, the low-mass stars, the pattern of enrichment produced among these stars might resemble that due to interior mixing. During their pre-main-sequence phase of evolution, the low-mass stars would have been highly convective, and any accreted material would be homogeneously spread throughout their interiors. One direct test of this scenario (though very difficult observationally) would be to measure the $^{12}C/^{13}C$ ratios of the 47 Tuc main-sequence stars, which are required to be very low.

Perhaps the least appealing, but arguably the most likely, explanation for the light element abundance patterns observed among globular clusters is that they are the product of a combination of both mixing and primordial processes. In this scenario, large N abundance inhomogeneities, and possibly moderate C and O inhomogeneities, would be fixed very early in a cluster's history. Then, during advanced stages of stellar evolution, circulation of the outer envelopes of cluster giants through an interior region of CN-, and possibly ON-, processing takes place. The efficiency of this mechanism is assumed to be metallicity dependent, with the metal-poorer stars undergoing the more extensive redistributions of surface C and O abundances. This evolutionary mechanism would be invoked to produce the more extreme luminosity-dependent carbon and oxygen abundance depletions found among the brightest cluster giants. However, the consequent C depletions are assumed to be insufficient to erase any CN variations that are the product of inhomogeneous primordial nitrogen enrichment.

A hybrid scenario could perhaps be formulated along these lines which would be consistent with all of the observations to date, including the fact that the C and N abundances of the halo field stars show evidence of dredge up, but little of the extensive CN variations seen in the clusters (such CN variations being postulated to be dependent upon a globular-cluster environment for the occurrence of primordial enrichment). Globular clusters of intermediate composition such as M4, NGC 6752, and M10 evince both CN bimodalities and weak signs of luminosity-dependent C depletions, while the metal-richer cluster 47 Tuc appears to have the same range in C and N abundances from the main-sequence to the RGB. These contrasting characteristics seem consistent with a hybrid scheme involving both primordial inhomogeneities and metallicity-dependent mixing. It is in the metal-richest clusters that stellar mixing would be least pronounced, such that the abundance inhomogeneities in clusters like 47 Tuc would largely reflect the primordial contributions. In clusters of intermediate metallicity such as M4 and NGC 6752, both primordial and mixing phenomena can be discerned. Mixing effects might dominate the inhomogeneities observed in the metal-poorest clusters such as M92 and M15. However, even in these clusters where mixing is perhaps most pronounced, and the metallicity is too low for any possible CN bimodality to be discerned, the C and N abundance data from Carbon, et al. [33] and Trefzger, et al. [34] are consistent with an underlying inhomogeneity not attributable to mixing.

An interesting experiment can be performed to probe what the CN distribution of the stars in M92 would look like if the metallicity of the cluster were sufficiently high that these bands could be observed. Model atmospheres appropriate to the effective temperatures and gravities of the least evolved RGB stars (those least likely to have had their C and N abundance ratios affected by mixing) in the Carbon et al. [33] M92 sample were computed, and synthetic spectra were generated using the observed [C/M] and [N/M] abundances for these stars [33], but an overall heavy element metallicity of [M/H] = –1.0 in order to ensure that the synthesized CN bands have a significant strength (for the metallicity of M92 used by Carbon et al., [M/H] = –2.0, the CN bands are very weak). The histogram of the resulting CN band strength indices is plotted in Figure 4, and is seen to exhibit a bimodality. Although an exercise such as this is far from rigorous, the similarities between Figure 4 and the CN distributions of other, more metal-rich clusters, suggests that the bimodal CN variations, and their underlying (primordial?) inhomogeneities in N (and perhaps C) abundances, are a property of most globular clusters.

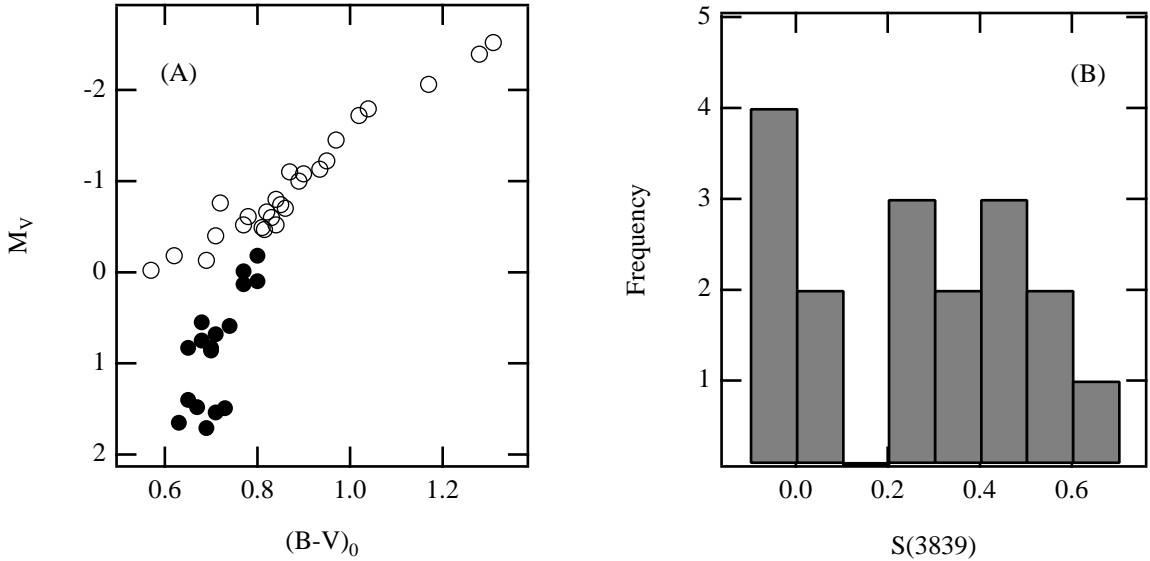

FIG. 4: (A) The M92 program stars observed by Carbon, et al. [33] and their positions in the color-magnitude diagram. Synthetic CN band strengths have been computed for the stars marked by closed symbols. (B) The distribution of CN band strengths assuming the C and N abundances of Carbon, et al., scaled to [M/H] = –1.0. The distribution appears similar to that seen in many of the more metal-rich clusters (see text).

4. Future Work

The use of diatomic molecules as probes of abundance patterns for stars within a given Galactic globular star cluster, and from one such cluster to another, has uncovered surprising and unanticipated results: there are clearly processes which have acted, or are acting, among the cluster stars to modify the surface composition of their light elements (namely, C, N, O, Al, and Na). An understanding of the nature of these mechanisms is important, not only for the study of stellar evolution, but also to permit the reliable use of globular clusters as observable probes of conditions in the early halo, and to set the chronology of the formation of the Galaxy (and indeed, the minimum age of the Universe itself).

One critical test of the role of internal processes is the determination of the total C + N + O abundances in the cluster stars: the CN- and ON-cycles of H burning do not modify C + N + O. Therefore, any star-to-star differences in this total must be the result of a mechanism external to the stars themselves. Oxygen abundances have recently been determined for a large number of stars in several clusters [52-57, 59] and observations of NH and CH features in these stars would be extremely valuable for the purpose of investigating whether C + N + O is constant within a cluster[5]. An example of such work is that of Brown, Wallerstein, and Oke [46] who find similar C + N + O abundances among five giants of differing CN band strengths in M13.

It is worth noting that the study of halo field stars also has a role to play in disentangling the origin of globular cluster abundance inhomogeneities. For example, it has been found that the pattern of [N/Fe] and [C/Fe] abundances among the halo field giants is different from that in the globular clusters [24], although the field giants do exhibit comparably low $^{12}C/^{13}C$ ratios to the cluster giants [40]. Extremely N-rich halo field stars are very rare, but those few that have been found show anomalies in the abundances of other elements as well. Spiesman [76] has found that the N-rich CN-strong subdwarf HD 25329 has a [Na/Fe] abundance ratio that is ~ 0.6 dex higher than those of three otherwise similar N-poor subdwarfs. Furthermore, Magain [77] finds enhanced Al lines in spectra of several N-rich halo dwarfs. Such correlated N-Na-Al abundance patterns among subdwarfs cannot be attributed to red giant nucleosynthesis of the type proposed by Langer, Hoffman, and Sneden [73], and beg the question as to whether primordial N-Na-Al correlations exist among Galactic globular cluster main-sequence stars.

---

[5] We note that CN and CH band strengths are technically sufficient for this purpose. However, errors in the determination of C abundances from CH will result in anticorrelated errors in the N abundances deduced from CN, thereby mimicking the effects of CN-processing.

The goals of research into the nature of abundance inhomogeneities in globular clusters are now seen to be twofold: to study element nucleosynthesis and mixing/dredge-up within the cluster red giants, and to characterize the "signature" of any underlying primordial abundance enhancements in the clusters. This second goal is necessary to permit identification of the nucleosynthetic process(es) that enriched the "primordial" material from which the cluster originated (*i.e.*, stellar winds from intermediate-mass stars, remnants of Type II supernovae, etc.). Moreover, constraints on the total mass of enriched material which must have been contributed by stars to the early intracluster environment is of value for addressing questions regarding the number of stars formed as a function of mass, as well as other properties of the early cluster environment and the general process of cluster and galaxy formation.

We wish to thank all of the researchers who have contributed to the study of globular cluster abundance inhomogeneities over the past two decades. Page limitations make it unfeasible to include a comprehensive list of references to all of their work, and we apologize to those whose published papers have not been cited here. JEH wishes to thank Dr. Herzberg for invaluable advice and encouragement early in his scientific career. We also thank the Cerro Tololo Inter-American Observatory for making the crucial 47 Tucanae observations possible. MMB acknowledges the support of the Texas Advanced Research Projects program during the writing of this review.